\documentclass[twocolumn,showpacs,amsmath,amssymb,pra]{revtex4}

 \usepackage{graphicx}
 \usepackage{bm}

\begin{document}

 \title{Medium effects on the van der Waals force}
 \author{M. S. Toma\v s}
 \email{tomas@thphys.irb.hr}
 \affiliation{Rudjer Bo\v skovi\' c Institute, P. O. B. 180,
 10002 Zagreb, Croatia}
 \date{\today}

 \begin{abstract}
We consider the van der Waals interaction between two ground-state
atoms embedded in adjacent semi-infinite magnetodielectric media,
with emphasis on medium effects on it. We demonstrate that, in
this case, at small atom-atom distances the van der Waals
interaction is screened by the surrounding media in the same way
as in an effective (single) medium. At larger atomic distances,
however, its dependence on the material parameters of the system
and the positions of the atoms is more complex. We also calculate
the Casimir-Polder potential of an atom A arising from a uniform
distribution of atoms B in the medium across the interface.
Comparison of this potential with the corresponding result deduced
from the Casimir force on a thin composite slab in front of a
composite semi-infinite medium, both obeying the Clausius-Mossotti
relation, suggests a hint on how to improve a well-known formula
for the van der Waals potential with respect to the local-field
effects.

\end{abstract}
 \preprint{IRB-TH-4/06}
 \maketitle

Owing to the gradually increasing role of the van der Waals (and
Casimir) forces with decreasing dimensions of the system on the
one side and rapid progress in miniaturization of modern
technologies on the other side, the van der Waals (atom-atom)
interaction in complex systems is an issue of great importance,
both fundamentally and practically. The van der Waals ineraction
in a confined space was usually considered assuming the atoms in
an empty region bounded by perfectly reflecting (conducting)
walls, e.g., in a planar cavity \cite{MN2,MN3} or in front of a
plate \cite{Pow,SPR}. Its properties near realistic boundaries
have been addressed only very recently \cite{Saf,Buh1}. However,
although it is known for quite some time that the surrounding
medium \cite{Dzy,Abr} has strong effects on the van der Waals
interaction, considerations of the combined medium and boundary
effects in realistic systems on it are very rare. Actually, so far
only Marcovitch and Diamant have addressed the van der Waals
interaction in a system of this kind (namely, a three-layer
dielectric system) and demonstrated its strong modification with
the material parameters of the system \cite{MaDi}. One of the
reasons for this is certainly the necessity of consideration of
the local-field effects on the atom-atom force in material
systems; an issue which so far has not been explicitly addressed.
In this work, we extend our previous consideration of the van der
Waals force in a magnetodielectric medium \cite{Tom06} to the case
when the atoms are embedded in different semi-infinite media, with
emphasis on medium effects on this force. Besides being of obvious
interest, e.g., in surface physics and related sciences, these
considerations also provide a hint on the local-field corrections
in the theory of the van der Waals interaction \cite{Tom06}.

\section{van der Walls interaction across an interface}
Consider two electrically polarizable atoms A i B embedded in an
inhomogeneous magnetodielectric system described by the
permittivity $\varepsilon({\bf r},\omega)$ and permeability
$\mu({\bf r},\omega)$. The van der Waals interaction energy
between the atoms is then given by
\begin{eqnarray}
\label{UAB} U_{AB}({\bf r}_A,{\bf r}_B)&= &-\frac{\hbar}{2\pi
c^4}\int_0^\infty d\xi
\xi^4\alpha_A(i\xi)\alpha_B(i\xi)\\
&&\times{\rm Tr}\left[\tensor{\bf G}({\bf r}_A,{\bf
r}_B;i\xi)\cdot\tensor{\bf G}({\bf r}_B,{\bf
r}_A;i\xi)\right],\nonumber
\end{eqnarray}
where $\alpha_{A(B)}(\omega)$ are the atomic vacuum
polarizabilities and $\tensor{\bf G}({\bf r},{\bf r'};\omega)$ is
the classical Green function for the system satisfying
 \begin{align}
 &\left[\nabla\times\frac{1}{\mu({\bf r},\omega)}\nabla\times-
 \varepsilon({\bf r},\omega)
 \frac{\omega^2}{c^2}\tensor{\bf I}\cdot\right]
 \tensor{\bf G}({\bf r},{\bf r'};\omega)\nonumber\\
 &\hspace{28ex}=4\pi\tensor{\bf I}
 \delta({\bf r}-{\bf r'}),
 \label{GF}
 \end{align}
 with the outgoing wave condition at infinity. This form for
 $U_{AB}({\bf r}_A,{\bf r}_B)$ was firstly obtained by Mahanty and
 Ninham  for two atoms in the free space \cite{MN1} and, relying on heuristic
 arguments, it was generally believed that, with the appropriate Green
 function, Eq. (\ref{UAB}) properly describes the van der Waals potential
 in inhomogeneous systems as well \cite{MN3,MaDi}. Very recently, this
 conjecture has been proved in various ways to be indeed correct,
 provided that the local-field effects can be neglected \cite{Saf,Buh1,RW}.

\begin{figure}[htb]
\begin{center}
 \resizebox{8cm}{!}{\includegraphics{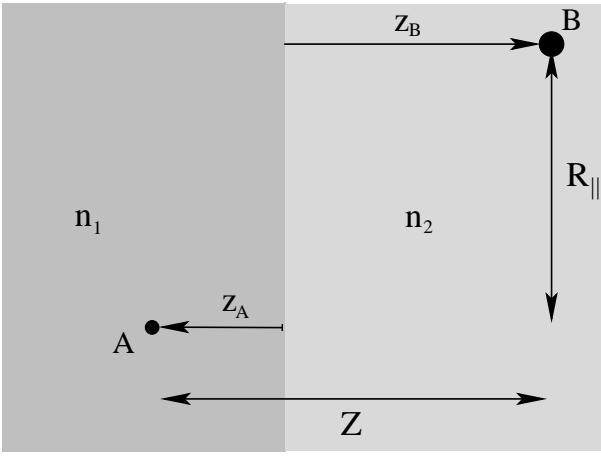}}
 \end{center}
 \caption{Two atoms interacting across an interface shown schematically.
 Media are described by (complex) refraction indexes
 $n_i(\omega)=\sqrt{\varepsilon_i(\omega)\mu_i(\omega)}$.
\label{AB}}
\end{figure}

Assuming that the atoms $A$ and $B$ are embedded, respectively, in
medium $1$ occupying the half-space $z<0$ and medium $2$ occupying
the half-space $z>0$, as depicted in Fig. \ref{AB}, we have
\cite{Tom95}

\[\tensor{\bf G}({\bf r}_A,{\bf r}_B;i\xi)=\int\frac{{\rm
d}^2{\bf k}}{(2\pi)^2}e^{i{\bf k}\cdot({\bf r}_{A\parallel}-{\bf
r}_{B\parallel})}\tensor{\bf G}({\bf k},i\xi;z_A,z_B),\]
\begin{eqnarray}
\tensor{\bf G}({\bf k},i\xi;z_A,z_B)&=&2\pi\frac{\mu_1}{\kappa_1}
\left[t^p_{12}\frac{i\kappa_1\hat{\bf k}+k\hat{\bf
z}}{k_1}\frac{i\kappa_2\hat{\bf k}+k\hat{\bf
z}}{k_2}\right.\nonumber\\
&&+\left.t^s_{12}\hat{\bf k}\times\hat{\bf z}\hat{\bf
k}\times\hat{\bf z}\right]e^{\kappa_1z_A-\kappa_2z_B}.
\label{G12}
\end{eqnarray}
Here $i\kappa_i$, with
\begin{equation}\label{kpa}
 \kappa_i=\sqrt{-k^2_i(i\xi)+k^2}=\sqrt{n^2_i(i\xi)\frac{\xi^2}{c^2}+k^2}
\end{equation}
is the perpendicular wave vector at the imaginary frequency in the
$i$th medium, whereas $r^q_{12}(i\xi,k)$ and
\begin{equation}\label{tr}
t^q_{12}(i\xi,k)=\sqrt{\frac{\gamma^q_{12}}{\gamma^s_{12}}}(1+r^q_{12})=
\sqrt{\frac{\gamma^q_{12}}{\gamma^s_{12}}}\frac{2\kappa_1}
{\kappa_1+\gamma^q_{12}\kappa_2},
\end{equation}
with $\gamma^p_{12}=\varepsilon_1/\varepsilon_2$ and
$\gamma^s_{12}=\mu_1/\mu_2$, are the Fresnel coefficients for the
$1-2$ interface.

Equations (\ref{UAB})-(\ref{tr}) provide a straightforward way for
calculating the van der Waals interaction energy between two atoms
in different media. Evidently, in this case, $U_{AB}$ is
anisotropic and depends not only on the distance between the
atoms, but also on their mutual orientation with respect to the
interface between the media. Clearly, the simplest situation for
consideration is when the atoms lie on a line perpendicular to the
interface, i.e., when ${\bf r}_{A\parallel}={\bf
r}_{B\parallel}=0$ and, to examine the medium effects on the van
der Waals interaction in the present system, we briefly consider
the small- and large-distance behavior of $U_{AB}$ in this
particular case.

Letting ${\bf r}_{A\parallel}={\bf r}_{B\parallel}=0$ in Eq.
(\ref{G12}) and performing the angular integration [${\bf
k}=k(\cos\varphi\hat{\bf x}+\sin\varphi\hat{\bf y})$], the Green
function takes the diagonal form
\begin{subequations}
\begin{eqnarray}
\tensor{\bf G}({\bf r}_A,{\bf
r}_B;i\xi)&=&G_\parallel(z_A,z_B;i\xi)(\hat{\bf x}\hat{\bf
x}+\hat{\bf y}\hat{\bf
y})\nonumber\\
&&+G_\perp(z_A,z_B;i\xi)\hat{\bf z}\hat{\bf z},\label{Gdia}
\end{eqnarray}
\begin{equation}
G_\parallel(z_A,z_B;i\xi)=\frac{\mu_1}{2}\int_0^\infty
\frac{dkk}{\kappa_1}(t^s_{12}-\kappa_1\kappa_2\frac{t^p_{12}}{k_1k_2})
e^{\kappa_1z_A-\kappa_2z_B},\label{Gpar}
\end{equation}
\begin{equation}
G_\perp(z_A,z_B;i\xi)=\mu_1\int_0^\infty
\frac{dkk^3}{\kappa_1}\frac{t^p_{12}}{k_1k_2}
e^{\kappa_1z_A-\kappa_2z_B}\label{Gper}.
\end{equation}
\end{subequations}
Using
\begin{equation}
\tensor{\bf G}({\bf r}_B,{\bf r}_A;i\xi)= \tensor{\bf G}^T({\bf
r}_A,{\bf r}_B;i\xi),
\end{equation}
we find that, in this case, the van der Waals potential is  given
by
\begin{eqnarray}
\label{UABdia}
U_{AB}({\bf r}_A,{\bf r}_B)&= &-\frac{\hbar}{2\pi
c^4}\int_0^\infty d\xi
\xi^4\alpha_A(i\xi)\alpha_B(i\xi)\\
&&\times\left[2G^2_\parallel(z_A,z_B;i\xi)
+G^2_\perp(z_A,z_B;i\xi)\right].\nonumber
\end{eqnarray}
Evidently, the integrals in Eqs. (\ref{Gpar}) and (\ref{Gper})
cannot be calculated analytically in the general case. For
$n_1=n_2\equiv n$, we obtain
\begin{subequations}
\begin{equation}
G_\parallel(z_A,z_B;i\xi)=\frac{\mu}{z}\left[1+ \frac{c}{n\xi
z}+(\frac{c}{n\xi z})^2\right]e^{-\frac{n\xi z}{c}},
\end{equation}
\begin{equation}
G_\perp(z_A,z_B;i\xi)=-\frac{2\mu}{z}\frac{c}{n\xi
z}\left(1+\frac{c}{n\xi z}\right)e^{-\frac{n\xi z}{c}},
\end{equation}
\end{subequations}
which, of course, in conjunction with Eq. (\ref{UABdia}) gives the
extension of the well-known result for the free-space van der
Waals potential \cite{MN1} to magnetodielectric media (see also
Refs. \cite{Saf,Tom06}).

The integral in Eq. (\ref{UABdia}) effectively extends up to a
frequency $\omega_{\rm max}$ correspondig to the largest
characteristic frequency of the system \cite{Saf}. Therefore,
owing to the presence of the exponential factors in Eqs.
(\ref{Gpar}) and (\ref{Gper}), the main contribution to
$G_\parallel(z_A,z_B;i\xi)$ and $G_\perp(z_A,z_B;i\xi)$ at small
atomic distances, $z_b-z_A\ll c/\omega_{\rm max}$, comes from the
large-$k$ waves. Letting $\kappa_i\rightarrow k$ in the integrands
, we obtain their nonretarded values
\begin{eqnarray}
G_\parallel(z_A,z_B;i\xi)&\simeq&\frac{c^2}{\xi^2}
\frac{2}{\varepsilon_1+\varepsilon_2}\frac{1}{Z^3},\\
G_\perp(z_A,z_B;i\xi)&\simeq&-\frac{c^2}{\xi^2}
\frac{2}{\varepsilon_1+\varepsilon_2}\frac{2}{Z^3},
\end{eqnarray}
where $Z=z_B-z_A$. With this inserted in Eq. (\ref{UABdia}), we
find that at small distances between the atoms
\begin{equation}
\label{UABs} U_{AB}({\bf r}_A,{\bf r}_B)=-\frac{3\hbar}{\pi Z^6}
\int_0^\infty d\xi
\frac{\alpha_A(i\xi)\alpha_B(i\xi)}{\bar{\varepsilon}^2(i\xi)},
\end{equation}
where $\bar\varepsilon=(\varepsilon_1+\varepsilon_2)/2$, i.e., the
same result as if the atoms were embedded in a single medium
\cite{Dzy,Saf,Tom06} with the dielectric function
$\bar\varepsilon(\omega)$.

At larger atom-atom distances, retardation of the electromagnetic
field starts to play a role and, owing to the different speed of
light in the two media, $U_{AB}$ is not a function of $Z$ any more
but rather a function of separate atomic coordinates $z_A$ and
$z_B$. To see this more clearly, we proceed in the standard way
\cite{Lif} and make the substitution $\kappa_1=n_1\xi p/c$ in Eqs.
(\ref{Gpar}) and (\ref{Gper}). We obtain
\begin{subequations}
\begin{eqnarray}
G_{\parallel(\perp)}(z_A,z_B;i\xi)&=&\mu_1\frac{n_1\xi}{c}\int_1^\infty
dpg_{\parallel(\perp)}(p,i\xi)\nonumber\\
&&\times e^{-\frac{n_1\xi}{c}(sz_B-pz_A)},\label{Gpp}
\end{eqnarray}
\begin{equation}
g_\parallel(p,i\xi)=
\frac{\mu_2p}{\mu_2p+\mu_1s}+p^2\frac{\varepsilon_1s}
{\varepsilon_2p+\varepsilon_1s},
\end{equation}
\begin{equation}
g_\perp(p,i\xi)=2(1-p^2)\frac{\varepsilon_1p}
{\varepsilon_2p+\varepsilon_1s},
\end{equation}
\end{subequations}
where $s(p,i\xi)=\sqrt{p^2-1+n_2^2/n_1^2}$. Inserting this into
Eq. (\ref{UABdia}) and changing the order of integrations, we have
\begin{align}
&U_{AB}({\bf r}_A,{\bf r}_B)= -\frac{\hbar}{2\pi c^6}\int_1^\infty
dp\int_1^\infty dp'\int_0^\infty d\xi\xi^6\alpha_A\alpha_B
\nonumber\\
&\hspace{1ex}\times\mu_1^2n_1^2[2g_\parallel(p,i\xi)
g_\parallel(p',i\xi)+g_\perp(p,i\xi) g_\perp(p',i\xi)]
\nonumber\\
&\hspace{21ex}\times e^{-\frac{n_1\xi}{c}[(s+s')z_B-(p+p')z_A)]}.
\end{align}
Now, for $z_A\omega_{\rm min}/c\gg 1$ and/or $z_B\omega_{\rm
min}/c\gg 1$, where $\omega_{\rm min}$ is the minimal
characteristic frequency of the atoms and the surrounding media
\cite{Saf}, the main contribution to the integral over $\xi$ comes
from the $\xi\simeq 0$ region. Approximating the
frequency-dependent quantities with their static values, the
integration becomes elementary and for the van der Waals potential
at large distances between the atoms we obtain
\begin{align}
\label{UABl} &U_{AB}({\bf r}_A,{\bf r}_B)=-\frac{360\hbar
c}{\pi}\frac{\alpha_A(0)\alpha_B(0)}{\varepsilon_1^2(0)n_1(0)}
\int_1^\infty dp\int_1^\infty dp'\nonumber\\
&\hspace{5ex}\times\frac{2g_\parallel(p,0)g_\parallel(p',0)+
g_\perp(p,0) g_\perp(p',0)}{[(s_0+s'_0)z_B-(p+p')z_A]^7},
\end{align}
where $s_0=\sqrt{p^2-1+n_2^2(0)/n_1^2(0)}$. As one may easily
verify, for a single medium ($n_2=n_1$) this result reduces to
\cite{Tom06,Saf}
\begin{equation}
\label{UABinf} U^{(1)}_{AB}({\bf r}_A,{\bf r}_B)= -\frac{23\hbar
c}{4\pi
}\frac{\alpha_A(0)\alpha_B(0)}{\varepsilon_1^2(0)n_1(0)Z^7}.
\end{equation}

\begin{figure}[htb]
\begin{center}
 \resizebox{8cm}{!}{\includegraphics{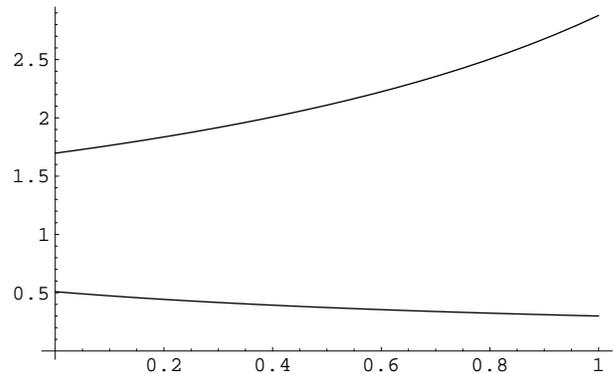}}
 \end{center}
 \caption{Relative van der Waals potential $U_{AB}/U^{(1)}_{AB}$ as a
 function of $z_B/Z$ for a large atom-atom distance $Z=z_B-z_A$. Media
 are assumed nonmagnetic ($\mu_1=\mu_2=1$). The upper and the lower curve
 correspond to $\varepsilon_2(0)=0.5\varepsilon_1(0)$ and
 $\varepsilon_2(0)=2\varepsilon_1(0)$, respectively.\label{Ularge}}
\end{figure}

Evidently, as before, there is a combined medium effect on
$U_{AB}$ at large atom-atom distances. This time, however, the
strength of the screening of the van der Waals interaction by each
medium is determined by the position of the atom embedded in it.
We illustrate this in Fig \ref{Ularge} where we have plotted the
ratio $U_{AB}/U^{(1)}_{AB}$ in a nonmagnetic system as a function
of $z_B$ for a fixed (large) distance $Z$ between the atoms and
for different values of $\varepsilon_2(0)/\varepsilon_1(0)$.
At $z_B=0$, except for its modification because of the field
transmission at the interface (described by the nominator in Eq.
(\ref{UABl})), the potential is screened entirely by medium $1$.
With increasing $z_B$, the screening of $U_{AB}$ by medium 1 is
gradually replaced by that of medium 2. Accordingly, since
$U^{(2)}_{AB}/U^{(1)}_{AB}=\varepsilon^{5/2}_1(0)/\varepsilon^{5/2}_2(0)$,
the relative potential $U_{AB}/U^{(1)}_{AB}$ increases for
$\varepsilon_2(0)<\varepsilon_1(0)$ and decreases for
$\varepsilon_2(0)>\varepsilon_1(0)$.

\section{Local-field corrections}

\begin{figure}[htb]
\begin{center}
 \resizebox{8cm}{!}{\includegraphics{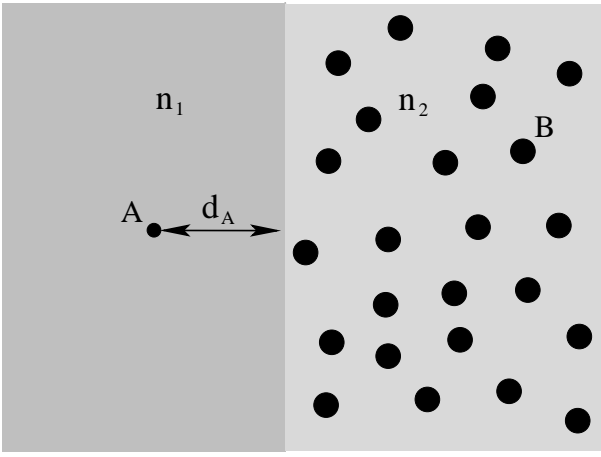}}
 \end{center}
 \caption{An atom interacting with a distribution of atoms across
 an interface.\label{A(B)}}
\end{figure}

As already mentioned, Eq. (\ref{UAB}) does not account for the
local-filed effects appearing in optically dense media. A hint on
how to improve this result with respect to the local-field effects
can be found by considering the Casimir-Polder potential as
implied by Eqs. (\ref{UAB}) and (\ref{G12}) in the case of a
(uniform) distribution of B atoms across the medium 2 (see Fig.
\ref{A(B)})
and by comparing it with the corresponding result deduced from the
Casimir force on a thin slab in front of a composite medium
obeying the Claussius-Mossotti equation \cite{Born}.

Assuming the atomic number density $N_B$ small enough, the
Casimir-Polder potential $U^{(B)}_A$ of the atom $A$ arising from
its interaction with atoms $B$ is obtained by pairwise summation
of the van der Waals potentials $U_{AB}$, i.e.,
\begin{equation}\label{Uint}
U^{(B)}_A({\bf r}_A)=\int_{z_B\geq 0}{\rm d}^3{\bf r}_B
U_{AB}({\bf r}_A,{\bf r}_B).
\end{equation}
Equations (\ref{UAB}) and (\ref{G12}) straightforwardly lead to
(see the Appendix)
\begin{align}\label{UA}
&U^{(B)}_A({\bf r}_A)=-\frac{N_B\hbar}{c^4}\int_0^\infty d\xi
\xi^4\mu_1\alpha_A\mu_2\alpha_B\int_0^\infty
\frac{dkk}{\kappa_2^2}\\
&\times \left[\left(\frac{2\kappa_1^2c^2}{n_1^2\xi^2}-1\right)
\left(\frac{2\kappa_2^2c^2}{n_2^2\xi^2}-1\right)t^p_{12}t^p_{21} +
t^s_{12}t^s_{21}\right]\frac{e^{2\kappa_1z_A}}{2\kappa_1},\nonumber
\end{align}
where the dependence of the quantities in the integrand on $i\xi$
and $(i\xi,k)$ is understood.

\begin{figure}[htb]
\begin{center}
 \resizebox{8cm}{!}{\includegraphics{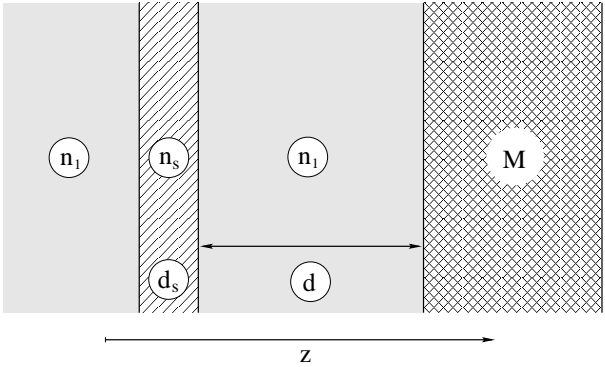}}
 \end{center}
 \caption{A slab in front of a mirror shown schematically. The
 (complex) refraction index of the slab is
 $n_s(\omega)=\sqrt{\varepsilon_s(\omega)\mu_s(\omega)}$ and that
 of the surrounding medium is $n_1(\omega)=\sqrt{\varepsilon_1(\omega)\mu_1(\omega)}$.
 The mirror is described by its reflection coefficients
 $R^q(\omega,k)$, with $k$ being the in-plane wave vector of a wave.
 The arrow indicates the direction of the force on the slab.\label{slab}}
\end{figure}

On the other hand, the potential $U^{(B)}_A$ can be deduced from
the Casimir force on a thin slab consisting of a layer of the
surrounding medium with a small number of foreign atoms embedded
in it in front of a composite medium \cite{Tom06}. We start from
the formula for the Casimir force on a slab ($s$) in a medium near
a mirror (as depicted in Fig. \ref{slab},) \cite{Tom02}
\begin{equation}
\label{f}
 f_s(d)=\frac{\hbar}{2\pi^2}\int_0^\infty
d\xi \int^\infty_0dkk\kappa_1
\sum_{q=p,s}\frac{r^qR^qe^{-2\kappa_1 d}} {1-r^qR^qe^{-2\kappa_1
d}},
\end{equation}
where $r^q(i\xi,k)$ are reflection coefficients of the
(symmetrically bounded) slab and $R^q(i\xi,k)$ are those of the
mirror. For a thin slab, so that $\kappa_sd_s\ll 1$ in the
relevant frequency range, we have
\begin{equation}
\label{rs} r^q(i\xi,k)=r^q_{1s}\frac{1-e^{-2\kappa_s
d_s}}{1-{r^q_{1s}}^2e^{-2\kappa_s d_s}}\simeq
2r^q_{1s}\kappa_sd_s.
\end{equation}
Assuming that the surrounding medium is a collection of
polarizable particles (atoms or molecules), the dielectric
function of the slab is given by the Clausius-Mossotti equation
\cite{Born} (the frequency dependence of $\varepsilon$'s and
$\alpha$'s is understood)
\begin{equation}
\frac{\varepsilon_s-1}{\varepsilon_s+2}=
\frac{4\pi}{3}(N_1\alpha_1+N_A\alpha_A)
=\frac{\varepsilon_1-1}{\varepsilon_1+2}+
\frac{4\pi}{3}N_A\alpha_A,
\end{equation}
where in the last line we have again used the Clausius-Mossotti
equation, this time for medium 1 alone. Accordingly, the
dielectric function of the slab can be written as
\begin{equation}
\label{es}
 \varepsilon_s=\varepsilon_1+4\pi N_A\tilde{\alpha}_A,
\end{equation}
where $\tilde{\alpha}_A$ is the effective polarizability of an $A$
atom given by
\begin{equation}\label{alef}
\tilde{\alpha}_A=\frac{\alpha_A (\frac{\varepsilon_1+2}{3})^2}
{1-\frac{4\pi}{3}N_A\alpha_A\frac{\varepsilon_1+2}{3}} \simeq
\alpha_A (\frac{\varepsilon_1+2}{3})^2,
\end{equation}
with the last line being valid when $N_A\alpha_A\ll 1$.

With Eq. ({\ref{es}), we have for small $N_A\tilde{\alpha}_A$
\begin{equation}\label{kps}
\kappa_s\simeq \kappa_1 (1+2\pi N_A\tilde{\alpha}_A\mu_1
\frac{\xi^2}{\kappa^2_1c^2}),
\end{equation}
so that the medium-slab reflection coefficients are to the first
order in $N_A\tilde{\alpha}_A$ given by
\begin{subequations}
\label{r1s}
\begin{equation}
r^p_{1s}=\frac{\varepsilon_s\kappa_1-\varepsilon_1\kappa_s}
{\varepsilon_s\kappa_1+\varepsilon_1\kappa_s} \simeq \frac{2\pi
N_A\tilde{\alpha}_A}{\varepsilon_1}(1-\frac{n^2_1\xi^2}{2\kappa_1^2c^2}),
\end{equation}
\begin{equation}
r^s_{1s}=\frac{\kappa_1-\kappa_s}{\kappa_1+\kappa_s} \simeq -\pi
N_A\tilde{\alpha}_A\mu_1\frac{\xi^2}{\kappa_1^2c^2}.
\end{equation}
\end{subequations}
Combining  Eqs.(\ref{kps}) and (\ref{r1s}) with Eq. (\ref{rs}) and
inserting these $r^q$'s into Eq. (\ref{f}), we find
\begin{equation}
f_s(d)=N_Ad_sf_A(d),
\end{equation}
where, with $d\equiv d_A$,
\begin{eqnarray}
 \label{fa}
f_A(d_A)&=&\frac{\hbar}{\pi c^2}\int_0^\infty d\xi\xi^2\mu_1
\tilde{\alpha}_A\int^\infty_0dkk e^{-2\kappa d_A}\nonumber\\
&&\times\left[\left(2\frac{\kappa^2_1c^2}{n_1^2\xi^2}-1\right)R^p
-R^s\right]
\end{eqnarray}
is the Casimir-Polder force on an atom \cite{Tom06}. We note that
this equation extends (in different directions) previous results
for the atom-mirror force in various circumstances
\cite{CP,Boy,Schw,Zhou} by accounting for the magnetic properties
of the media (see also Refs. \cite{Buh,Buh2,Tom052}) and including
the local-field corrections within the Lorentz model \cite{Lor}
for the local field.

Equation (\ref{fa}) enables one to calculate the force on the atom
$A$ due to the uniform distribution of atoms $B$ in a
magnetodielectric medium $2$. Assuming that the mirror is a
mixture of type $2$ (electrically) polarizable particles and type
$B$ atoms, its dielectric function $\varepsilon_m$ and the
perpendicular wave vector $\kappa_m$ inside it are given by Eqs.
(\ref{es})-(\ref{kps}), with $\{s,1,A\}\rightarrow \{m,2,B\}$.
Accordingly, for the reflection coefficients of the mirror we find
to the first order in $N_B\tilde{\alpha}_B$
\begin{subequations}
\begin{eqnarray}
R^p&=&\frac{\varepsilon_m\kappa_1-\varepsilon_1\kappa_m}
{\varepsilon_m\kappa_1+\varepsilon_1\kappa_m}=r^p_{12}\nonumber\\
&&+t^p_{12}t^p_{21}\frac{\pi
N_B\tilde{\alpha}_B\mu_2\xi^2}{\kappa_2^2c^2}(\frac{2\kappa_2c^2}{n_2^2\xi^2}-1),
\end{eqnarray}
\begin{equation}
R^s=\frac{\mu_m\kappa_1-\mu_1\kappa_m}
{\mu_m\kappa_1+\mu_1\kappa_m}=r^s_{12}-t^s_{12}t^s_{21}\frac{\pi
N_B\tilde{\alpha}_B\mu_2\xi^2}{\kappa_2^2c^2},
\end{equation}
\end{subequations}
where the single-interface Fresnel coefficients $r^q_{12}$ and
$t^q_{12}$ are given by Eq. (\ref{tr}). Inserting this into Eq.
(\ref{fa}), we find for the Casimir-Polder force near such a
composite mirror
\begin{equation}
f_A(d_A)=f^{(2)}_A(d_A)+f^{(B)}_A(d_A),
\end{equation}
where $f^{(2)}_A(d_A)$ is the Casimir-Polder force of the atom in
the vicinity of medium $2$ alone [given by Eq. (\ref{fa}), with
$R^q\rightarrow r^q_{12}$] and $f^{(B)}_A(d_A)$ is the force on
the atom due to the uniform distribution of $B$ atoms across
medium $2$. As seen, this latter force coincides precisely with
the Casimir-Polder force obtained from Eq. (\ref{UA}) (note that
$z_A=-d_A$)
\begin{equation}
{\bf f}^{(B)}_A(z_A)=-\nabla_AU^{(B)}_A({\bf r}_A),
\end{equation}
provided that we let
\begin{equation}
\alpha_{A(B)}(i\xi)\rightarrow\tilde{\alpha}_{A(B)}(i\xi)
\simeq\alpha_{A(B)}(i\xi)\left[\frac{\varepsilon_{1(2)}(i\xi)+2}{3}\right]^2.
\end{equation}
This suggests that, with the above replacement, Eq. (\ref{UAB})
can also be used to describe the atom-atom interaction in
optically dense media where the local-field effects cannot be
neglected.

\section{Summary}
In this work we have presented basic equations for consideration
of the van der Waals interaction between two ground-state atoms
embedded in adjacent semi-infinite magnetodielectric media and
obtained a few results concerning the medium effects on this
interaction. By considering a simple configuration, we have
demonstrated that the atom-atom interaction in this system is at
small distances screened by the surrounding media in the same way
as in an effective (single) medium. At larger atomic distances,
however, its dependence on the material parameters of the system
and the positions of the atoms is more complex. We have also
calculated the Casimir-Polder potential of an atom A arising from
a collection of atoms B uniformly distributed in the medium across
the interface. Comparison of this potential with the corresponding
result deduced from the Casimir force on a thin composite slab in
front of a composite semi-infinite medium, both obeying the
Clausius-Mossotti relation, suggests that Eq. (\ref{UAB}) can be
adopted to describe the van der Waals potential in optically dense
media as well, provided that the atomic polarizabilities are
replaced by the effective ones.

\acknowledgments
 This work was supported in part by the Ministry of Science and
 Technology of the Republic of Croatia under contract No. 0098001.

\appendix*
\section{}
Combining Eqs. (\ref{UAB}) and (\ref{Uint}), we have
\begin{align}\label{A1}
&U^{(B)}_{A}({\bf r}_A)= -\frac{N_B\hbar}{2\pi c^4}\int_0^\infty
d\xi \xi^4\alpha_A\alpha_B\\
&\hspace{4ex}\times\int_{z_B\geq 0}{\rm d}^3{\bf r}_B {\rm
Tr}\left[\tensor{\bf G}({\bf r}_A,{\bf r}_B;i\xi) \cdot\tensor{\bf
G}({\bf r}_B,{\bf r}_A;i\xi)\right].\nonumber
\end{align}
Noting that \cite{Tom95}

\begin{equation}
\tensor{\bf G}({\bf r}_B,{\bf r}_A;i\xi)=\int\frac{{\rm d}^2{\bf
k}}{(2\pi)^2}e^{i{\bf k}\cdot({\bf r}_{B\parallel}-{\bf
r}_{A\parallel})}\tensor{\bf G}^T(-{\bf k},i\xi;z_A,z_B)
\end{equation}
and using Eq. (\ref{G12}), we find that the space integral in Eq.
(\ref{A1}) is equal
\begin{widetext}
\begin{equation}
\int\frac{{\rm d}^2{\bf k}}{(2\pi)^2}\int_0^\infty{\rm d}z_b {\rm
Tr}\left[\tensor{\bf G}({\bf k},i\xi;z_A,z_B)\cdot\tensor{\bf
G}^T(-{\bf k},i\xi;z_A,z_B)\right] =\int{\rm d}^2{\bf
k}(\frac{\mu_1}{\kappa_1})^2\left[(t^p_{12})^2\frac{\kappa_1^2+k^2}{k_1^2}
\frac{\kappa_2^2+k^2}{k_2^2}+(t^s_{12})^2\right]
\frac{e^{2\kappa_1z_A}}{2\kappa_2}.
\end{equation}
\end{widetext}}
Noting that $(\mu_1/\kappa_1)t^q_{12}=(\mu_2/\kappa_2)t^q_{21}$
and using Eq. (\ref{kpa}), we arrive at Eq. (\ref{UA}).

\end{document}